%% file: paper.tex
\documentclass{article}

\usepackage{arxiv}

\usepackage[utf8]{inputenc} 
\usepackage[T1]{fontenc}    
\usepackage{hyperref}       
\usepackage{url}            
\usepackage{booktabs}       
\usepackage{amsmath}        
\usepackage{amsfonts}       
\usepackage{nicefrac}       
\usepackage{microtype}      
\usepackage{cleveref}       
\usepackage{graphicx}
\usepackage{natbib}
\usepackage{doi}
\usepackage{xspace}
\usepackage{listings}       
\usepackage{xcolor}         
\usepackage{subcaption}     
\usepackage{multirow}       
\usepackage{tikz}           
\usetikzlibrary{arrows.meta,positioning,shapes.geometric,fit}

\newcommand{\code}[1]{\mbox{\texttt{#1}}}
\newcommand{\rosetta}{\mbox{LLM-Rosetta}\xspace}
\newcommand{\ir}{\mbox{IR}\xspace}

\lstdefinestyle{pythonstyle}{
    language=Python,
    basicstyle=\ttfamily\small,
    keywordstyle=\color{blue},
    commentstyle=\color{gray},
    stringstyle=\color{red!60!black},
    breaklines=true,
    frame=single,
    framesep=3pt,
    numbers=left,
    numberstyle=\tiny\color{gray},
    xleftmargin=2em,
}

\title{\rosetta: A Hub-and-Spoke Intermediate Representation\\for Cross-Provider LLM API Translation}

\newif\ifuniqueAffiliation
\uniqueAffiliationtrue

\ifuniqueAffiliation
\author{
\href{https://orcid.org/0000-0001-8353-0821}{\includegraphics[scale=0.06]{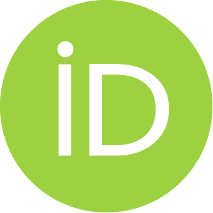}\hspace{1mm}Peng Ding} \\
University of Chicago \\
\texttt{dingpeng@uchicago.edu}
}
\fi

\renewcommand{\shorttitle}{LLM-Rosetta}

\hypersetup{
pdftitle={LLM-Rosetta: A Hub-and-Spoke Intermediate Representation for Cross-Provider LLM API Translation},
pdfsubject={Computer Science - Software Engineering},
pdfauthor={Peng Ding},
pdfkeywords={LLM, API translation, intermediate representation, interoperability, streaming, multi-provider},
}

\begin{document}
\maketitle

\begin{abstract}
The rapid proliferation of Large Language Model (LLM) providers---each exposing proprietary API formats---has created a fragmented ecosystem where applications become tightly coupled to individual vendors. Switching or bridging providers requires $O(N^2)$ bilateral adapters, impeding portability and multi-provider architectures. We observe that despite substantial syntactic divergence, the major LLM APIs share a common semantic core: the practical challenge is the \emph{combinatorial surface} of syntactic variations, not deep semantic incompatibility. Based on this finding, we present \rosetta, an open-source translation framework built on a \emph{hub-and-spoke} Intermediate Representation (\ir) that captures the shared semantic core---messages, content parts, tool calls, reasoning traces, and generation controls---in a 9-type content model and 10-type stream event schema. A modular \emph{Ops-composition} converter architecture enables each API standard to be added independently. \rosetta supports bidirectional conversion (provider$\leftrightarrow$\ir$\leftrightarrow$provider) for both request and response payloads, including chunk-level streaming with stateful context management. We implement converters for four API standards (OpenAI Chat Completions, OpenAI Responses, Anthropic Messages, and Google GenAI), covering the vast majority of commercial providers. Empirical evaluation demonstrates lossless round-trip fidelity, correct streaming behavior, and sub-100\,\textmu s conversion overhead---competitive with LiteLLM's single-pass approach while providing bidirectionality and provider neutrality. \rosetta passes the Open Responses compliance suite and is deployed in production at Argonne National Laboratory. Code is available at \url{https://github.com/Oaklight/llm-rosetta}.
\end{abstract}


\input{sections/intro}

\input{sections/related_work}

\input{sections/design}

\input{sections/implementation}

\input{sections/evaluation}

\input{sections/discussion}

\input{sections/conclusion}

\section*{Acknowledgments}

Large language models were used to assist with proofreading and language editing. The author takes full responsibility for all content.

\bibliographystyle{unsrtnat}
\bibliography{references}

\end{document}

%% file: sections/intro.tex
\section{Introduction}
\label{sec:intro}

Large Language Models (LLMs)~\citep{achiam2023gpt4,team2023gemini} are increasingly accessed through cloud-hosted APIs, yet the ecosystem lacks a universal wire format.
OpenAI's Chat Completions~\citep{openai2024chatapi}, Anthropic's Messages~\citep{anthropic2024messages}, Google's Generative AI~\citep{google2024geminiapi}, and the newer OpenAI Responses API~\citep{openai2025responses} each define their own schemas for messages, tool calls, streaming, and generation controls.
This divergence forces application developers to write provider-specific glue code, and organizations evaluating multiple models face a combinatorial integration burden.

\paragraph{The $O(N^2)$ problem.}
Given $N$ providers, na\"ive pairwise translation requires $\binom{N}{2}$ bilateral adapters.
Each adapter must handle request construction, response parsing, streaming event mapping, and tool-call serialization---all of which drift as providers evolve their APIs.
In practice, most projects either lock into a single vendor or adopt heavyweight SDK wrappers that hide---but do not solve---the underlying format mismatch.

\paragraph{Hub-and-spoke as $O(N)$ solution.}
The hub-and-spoke pattern---routing all conversions through a single Intermediate Representation (\ir)---is well established in compiler infrastructure~\citep{lattner2004llvm} and data interchange~\citep{arrow2016}, where it reduces $M \times N$ adapters to $M + N$.
LLM API translation is a structurally simpler domain (dict-to-dict mapping rather than semantic-preserving program transformation), but the same combinatorial argument applies: with $N$ providers and multiple feature dimensions, pairwise adapters grow quadratically.
We apply the hub-and-spoke principle to this domain.

\paragraph{Contributions.}
We present \rosetta, an open-source framework that introduces:
\begin{enumerate}
    \item An \textbf{empirical characterization of API divergence}: despite substantial syntactic differences, the four major LLM providers share a common semantic core that can be captured by a 9-type content model and 10-type stream event schema. The practical difficulty lies not in deep semantic gaps but in the \emph{combinatorial surface} of syntactic variations across providers and feature dimensions (\cref{sec:design:ir}).
    \item A \textbf{typed Intermediate Representation} (\ir) and \textbf{Ops-composition converter architecture} that exploit this observation: since divergence is predominantly syntactic, a provider-neutral IR can faithfully represent all four formats, and each provider adapter can be assembled from four orthogonal operations modules (content, message, tool, config) with effort independent of existing providers (\cref{sec:design}).
    \item \textbf{Bidirectional conversion with streaming support}: request and response payloads are translated in both directions (provider$\leftrightarrow$\ir$\leftrightarrow$provider), and chunk-level streaming is handled through ten typed event kinds with stateful context management (\cref{sec:impl}).
    \item An \textbf{empirical evaluation} demonstrating lossless round-trip fidelity, correct streaming behavior, and sub-millisecond conversion overhead, validated by 1{,}364 tests and the Open Responses compliance suite, and deployed in production at Argonne National Laboratory (\cref{sec:eval}).
\end{enumerate}

\rosetta currently supports four API standards---OpenAI Chat Completions, OpenAI Responses, Anthropic Messages, and Google Generative AI---which collectively cover the vast majority of commercial LLM providers, as most adopt one of these wire formats.
The code is released under the MIT license at \url{https://github.com/Oaklight/llm-rosetta}.

\paragraph{Paper organization.}
\Cref{sec:related} surveys related work.
\Cref{sec:design} presents the \ir design and converter architecture.
\Cref{sec:impl} describes the implementation, including streaming and the gateway proxy.
\Cref{sec:eval} evaluates round-trip fidelity, streaming correctness, and performance overhead.
\Cref{sec:discussion} discusses limitations and future directions.
\Cref{sec:conclusion} concludes.

%% file: sections/related_work.tex
\section{Related Work}
\label{sec:related}

\subsection{LLM API Ecosystem}

The landscape of LLM APIs has evolved rapidly since the release of GPT-3~\citep{brown2020language}.
OpenAI's Chat Completions API~\citep{openai2024chatapi} established an early de facto standard based on role-tagged messages (\code{system}, \code{user}, \code{assistant}), but subsequent providers diverged.
Anthropic's Messages API~\citep{anthropic2024messages} introduced a separate \code{system} parameter and block-typed content arrays.
Google's Generative AI API~\citep{google2024geminiapi} adopted a \code{contents}/\code{parts} schema with \code{user}/\code{model} roles.
OpenAI itself introduced the Responses API~\citep{openai2025responses}, replacing the chat message paradigm with an \emph{items-based} model where tool calls, reasoning, and text output are sibling items rather than nested content parts.

This fragmentation extends to tool calling~\citep{schick2023toolformer,patil2023gorilla,qin2023toolllm} (function definitions and invocation formats), streaming (SSE event schemas), multi-modal content~\citep{liu2024visual} (image, audio, file encoding), and generation controls (temperature, top-p, reasoning budgets~\citep{wei2022chain}).
\Cref{tab:api_divergence} summarizes key differences across providers.

\begin{table}[t]
\centering
\caption{API format divergence across major LLM providers.}
\label{tab:api_divergence}
\small
\begin{tabular}{@{}lllll@{}}
\toprule
\textbf{Aspect} & \textbf{OpenAI Chat} & \textbf{Anthropic} & \textbf{Google GenAI} & \textbf{OpenAI Responses} \\
\midrule
Message unit & \code{messages[]} & \code{messages[]} & \code{contents[]} & \code{input[]} items \\
System prompt & role in array & top-level param & role in array & \code{instructions} \\
Content model & string or parts & block array & \code{parts[]} & items with \code{type} \\
Tool calls & \code{tool\_calls[]} & content block & \code{functionCall} & item with \code{type} \\
Streaming & delta chunks & event types & candidate deltas & response events \\
Reasoning & --- & \code{thinking} & \code{thought} & \code{reasoning} item \\
\bottomrule
\end{tabular}
\end{table}

\subsection{SDK Wrappers and Abstraction Layers}

Several projects attempt to unify LLM access at different levels of abstraction.

\textbf{Multi-provider frameworks.}
LangChain~\citep{langchain2024} and Microsoft's Semantic Kernel~\citep{semantickernel2024} are the two most widely adopted multi-provider LLM frameworks.
Both provide high-level abstractions (chains, agents, planners) that internally dispatch to provider-specific SDK clients.
Their multi-provider support operates at the \emph{application} level: each provider integration is a separate adapter class that maps framework-level abstractions to native API calls.
This design prioritizes developer ergonomics for building LLM applications but does not expose a reusable, format-level translation layer---cross-provider conversion of raw API payloads is not a supported use case.

\textbf{SDK-level proxies.}
LiteLLM~\citep{litellm2024} provides an OpenAI-compatible proxy that translates requests at the SDK level, mapping all providers into the OpenAI Chat Completions format.
While widely adopted, this approach uses a single provider's schema as the lingua franca, which loses provider-specific features (e.g., Anthropic's cache control, Google's grounding metadata) and cannot represent constructs that the target format lacks.
AI Gateway~\citep{portkey2024aigateway} and similar commercial proxies route requests to multiple providers but typically rely on the OpenAI Chat format as the canonical schema, inheriting the same representational limitations.

\textbf{Specification and protocol efforts.}
The OpenRouter~\citep{openrouter2024} service aggregates providers behind a unified API, and the Open Responses~\citep{openresponses2025} initiative proposes the OpenAI Responses API format as an open standard adopted by multiple inference providers.
At a different layer, the Model Context Protocol (MCP)~\citep{mcp2024} standardizes how LLM applications discover and invoke tools, but does not address the request/response format translation between providers that \rosetta targets.

\subsection{Compiler Intermediate Representations}

The hub-and-spoke pattern is well established in compiler design.
LLVM's IR~\citep{lattner2004llvm} decouples source languages from target architectures, reducing adapter complexity from $O(M \times N)$ to $O(M + N)$.
Apache Arrow~\citep{arrow2016} applies the same principle to columnar data interchange between analytics systems.
Protocol Buffers~\citep{protobuf2008} and Apache Thrift~\citep{slee2007thrift} serve as language-neutral serialization IRs.

\rosetta adapts this hub-and-spoke strategy to the LLM API domain, which is structurally simpler than compiler IR (dict-to-dict mapping rather than semantic-preserving program transformation) but faces the same combinatorial cost: a typed \ir captures the semantic union of provider formats, while per-provider converters serve as frontends and backends.

\subsection{Positioning of \rosetta}

Unlike application frameworks (LangChain, Semantic Kernel) that abstract over providers at the application level, \rosetta operates at the \emph{API format} level, enabling raw payload translation without requiring adoption of a specific application framework.
Unlike SDK wrappers (LiteLLM) that privilege one provider's format, \rosetta defines a \emph{neutral} \ir designed from the ground up for cross-provider translation.
Unlike gateway proxies that focus on routing, \rosetta provides bidirectional, field-level conversion with explicit metadata preservation for lossless round-trips.
Unlike specification efforts (Open Responses), \rosetta is a runtime translation engine that can bridge existing, incompatible APIs without requiring providers to change their formats.
These distinctions do not imply that \rosetta supersedes existing tools.
LiteLLM's single-provider lingua franca is pragmatically effective for the common case of forwarding requests to diverse backends, and its ecosystem---100+ providers, built-in rate limiting, caching, observability, and an active community of over 15{,}000 GitHub stars---makes it the more practical choice for applications that do not require bidirectional or cross-provider translation.
LangChain and Semantic Kernel provide application-level abstractions (chains, agents, planners) that \rosetta does not attempt to replicate.
\rosetta is complementary: it can serve as the format-translation engine \emph{within} such frameworks or gateways, handling the low-level payload conversion that they currently implement ad hoc.
\Cref{tab:comparison} summarizes the positioning.

\begin{table}[t]
\centering
\caption{Comparison of \rosetta with related approaches. \texttimes{} = not supported; n/a = not applicable (specification, not implementation).}
\label{tab:comparison}
\small
\begin{tabular}{@{}lccccc@{}}
\toprule
\textbf{Feature} & \textbf{LangChain} & \textbf{LiteLLM} & \textbf{AI Gw.} & \textbf{Open R.} & \textbf{\rosetta} \\
\midrule
Provider-neutral IR      & \texttimes & \texttimes & \texttimes & n/a        & \checkmark \\
Bidir.\ format conv.     & \texttimes & \texttimes & \texttimes & n/a        & \checkmark \\
Lossless round-trip      & \texttimes & \texttimes & \texttimes & n/a        & \checkmark \\
Streaming support        & \checkmark & \checkmark & \checkmark & \checkmark & \checkmark \\
App framework            & \checkmark & \texttimes & \texttimes & \texttimes & \texttimes \\
Providers ($\ge$50)      & \checkmark & \checkmark & \checkmark & \texttimes & \texttimes \\
\bottomrule
\end{tabular}
\end{table}

%% file: sections/design.tex
\section{Design}
\label{sec:design}

This section presents the design of \rosetta's Intermediate Representation and converter architecture.
We first state the design goals (\cref{sec:design:goals}), then describe the \ir schema (\cref{sec:design:ir}), and finally introduce the Ops-composition pattern that structures each provider converter (\cref{sec:design:ops}).

\subsection{Design Goals}
\label{sec:design:goals}

\begin{enumerate}
    \item \textbf{Semantic completeness.} The \ir must be expressive enough to represent any construct found in supported providers---messages, multi-modal content, tool definitions and invocations, reasoning traces, generation controls, and streaming events---without loss of application-relevant information.
    \item \textbf{Provider neutrality.} No single provider's format should be privileged. The \ir is designed from the union of all supported schemas, not as an extension of any one.
    \item \textbf{Bidirectional fidelity.} Conversion must work in both directions (provider$\to$\ir and \ir$\to$provider) so that \rosetta can serve as both a request translator and a response translator.
    We define \emph{lossless round-trip} as follows: let $\mathit{to}_{A}$ and $\mathit{from}_{A}$ denote the to-IR and from-IR converters for provider~$A$, and let $\equiv_s$ denote structural equality (identical JSON trees modulo key ordering and insignificant whitespace). A round-trip is \emph{lossless in preserve mode} when $\mathit{from}_{A}(\mathit{to}_{A}(x)) \equiv_s x$ for every valid provider payload~$x$. In \emph{strip mode}, provider-specific metadata fields (e.g., \code{cache\_control}, \code{thought\_signature}) are intentionally discarded, so the property weakens to semantic equivalence: all application-relevant fields (message roles, content, tool definitions, generation parameters) are preserved, while provider-internal annotations may be dropped.
    \item \textbf{Incremental extensibility.} Adding a new provider should require implementing only the provider-specific converter without modifying the \ir schema or existing converters.
    \item \textbf{Streaming compatibility.} The design must support chunk-level streaming translation, not just batch request/response conversion.
\end{enumerate}

\subsection{Intermediate Representation}
\label{sec:design:ir}

\begin{figure}[t]
\centering
\includegraphics[width=\linewidth]{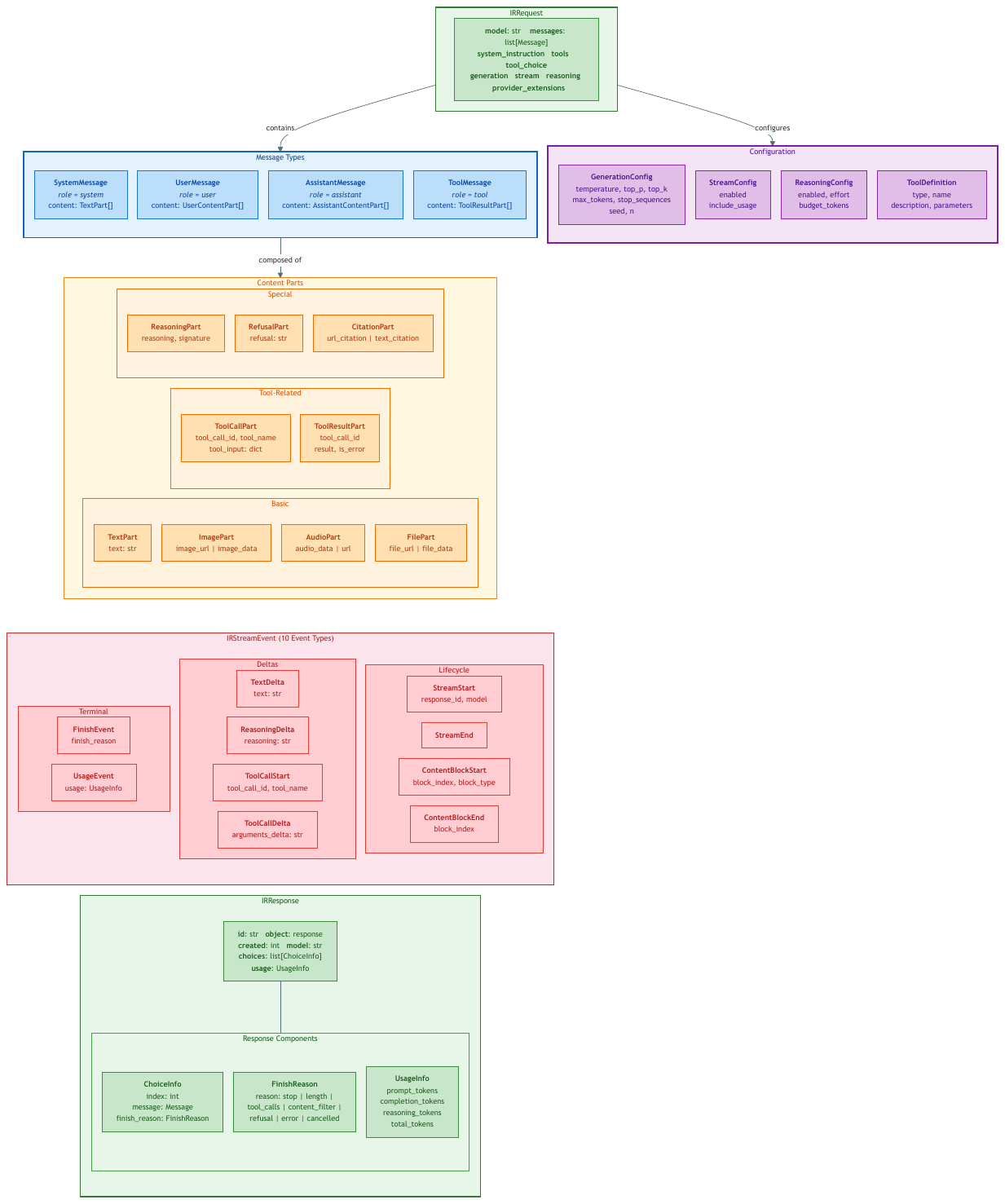}
\caption{Overview of the \ir schema. Arrows indicate containment relationships.}
\label{fig:ir_schema}
\end{figure}

The \ir is defined as a set of typed data structures organized into eight modules: content parts, messages, tools, generation configuration, requests, responses, stream events, and extension types.
\Cref{fig:ir_schema} provides an overview.

\subsubsection{Content Parts}

Content parts are the atomic units of message content.
The \ir defines the following part types:

\begin{itemize}
    \item \textbf{TextPart}: Plain text content.
    \item \textbf{ImagePart}: Image data (inline base64 or URL reference) with optional detail level.
    \item \textbf{AudioPart}: Audio data with media type.
    \item \textbf{FilePart}: Arbitrary file attachments.
    \item \textbf{ToolCallPart}: A tool invocation with call ID, tool name, and JSON input.
    \item \textbf{ToolResultPart}: The result of a tool invocation, linked by call ID.
    \item \textbf{ReasoningPart}: Chain-of-thought or ``thinking'' content, with optional signature for caching.
    \item \textbf{RefusalPart}: Model refusal with reason text.
    \item \textbf{CitationPart}: URL or text citations attached to generated content.
\end{itemize}

Each part carries a \code{type} discriminator and an optional \code{provider\_metadata} field for round-trip preservation of provider-specific attributes.

\subsubsection{Messages}

Messages are role-tagged containers of content parts:
\begin{itemize}
    \item \textbf{SystemMessage}: System-level instructions (role = \code{system}).
    \item \textbf{UserMessage}: User input, including text and images (role = \code{user}).
    \item \textbf{AssistantMessage}: Model output, including text, tool calls, and reasoning (role = \code{assistant}).
    \item \textbf{ToolMessage}: Tool execution results (role = \code{tool}).
\end{itemize}

Each message carries a \code{MessageMetadata} record with optional fields for message ID, timestamp, streaming state, and a \code{custom} dictionary for converter-specific round-trip data.

\subsubsection{Tool Definitions}

A \code{ToolDefinition} specifies a callable tool with:
\begin{itemize}
    \item \code{name}: Unique identifier.
    \item \code{description}: Natural-language description for the model.
    \item \code{parameters}: JSON Schema object defining the input shape.
    \item \code{type}: Tool category (\code{function} or \code{mcp}).
\end{itemize}

\code{ToolChoice} controls tool selection behavior with modes \code{none}, \code{auto}, \code{any}, and \code{tool} (force a specific tool).
\code{ToolCallConfig} provides additional controls such as disabling parallel tool calls.

\subsubsection{Generation Configuration}

\code{GenerationConfig} captures sampling and decoding parameters: temperature, top-p, top-k, max tokens, stop sequences, frequency/presence penalties, logit biases, seed, and logprobs settings.
\code{ReasoningConfig} controls chain-of-thought behavior (enabled, effort level, budget tokens).
\code{StreamConfig} and \code{ResponseFormatConfig} handle streaming and structured output settings.

\subsubsection{Request and Response}

\code{IRRequest} has two required fields---\code{model} and \code{messages}---and optional fields for system instruction, tools, tool choice, generation config, response format, streaming, reasoning, caching, and a \code{provider\_extensions} bag for rare provider-specific parameters that do not warrant first-class \ir fields.

\code{IRResponse} contains an ID, timestamp, model identifier, a list of \code{ChoiceInfo} (each wrapping a message and finish reason), and optional usage statistics (\code{UsageInfo} with prompt, completion, reasoning, and cache token counts).

\subsection{Ops-Composition Architecture}
\label{sec:design:ops}

Rather than implementing each converter as a monolithic class, \rosetta factors conversion logic into four orthogonal \emph{Ops} modules.
A monolithic converter intermingles content-level concerns (e.g., base64 image encoding) with request-level concerns (e.g., parameter mapping), causing cross-cutting logic like JSON Schema sanitization to be duplicated across all providers.
An alternative flat-parameter approach---mapping all provider fields through a single unified body---conflates semantic differences (what a field \emph{means}) with mechanical differences (how it is \emph{serialized}), making the converter brittle when providers share structure but differ in semantics.
The domain-factored Ops pattern isolates genuinely orthogonal concerns:

\begin{enumerate}
    \item \textbf{ContentOps}: Converts individual content parts (text, images, tool calls, reasoning, citations) between provider format and \ir.
    \item \textbf{MessageOps}: Converts message sequences, handling role mapping, system prompt extraction, and multi-turn conversation structure. Delegates to ContentOps for part-level conversion.
    \item \textbf{ToolOps}: Converts tool definitions and tool choice configurations.
    \item \textbf{ConfigOps}: Converts generation parameters, reasoning settings, response format, and caching configuration.
\end{enumerate}

A base class defines the abstract interface for each Ops module.
A concrete converter is assembled by specifying four Ops implementations:

\begin{lstlisting}[style=pythonstyle,caption={Ops-composition pattern.},label={lst:ops}]
class AnthropicConverter(BaseConverter):
    content_ops_class = AnthropicContentOps
    message_ops_class = AnthropicMessageOps
    tool_ops_class    = AnthropicToolOps
    config_ops_class  = AnthropicConfigOps
\end{lstlisting}

This design provides three benefits:
\begin{itemize}
    \item \textbf{Separation of concerns}: Content-level quirks (e.g., Google's \code{parts} nesting) are isolated from message-level concerns (e.g., Anthropic's separate system parameter).
    \item \textbf{Reuse}: Providers sharing a sub-format can reuse Ops modules. For example, OpenAI Responses reuses aspects of OpenAI Chat's content ops.
    \item \textbf{Testability}: Each Ops module can be unit-tested in isolation against known input/output pairs.
\end{itemize}

\begin{figure}[t]
\centering
\includegraphics[width=\linewidth]{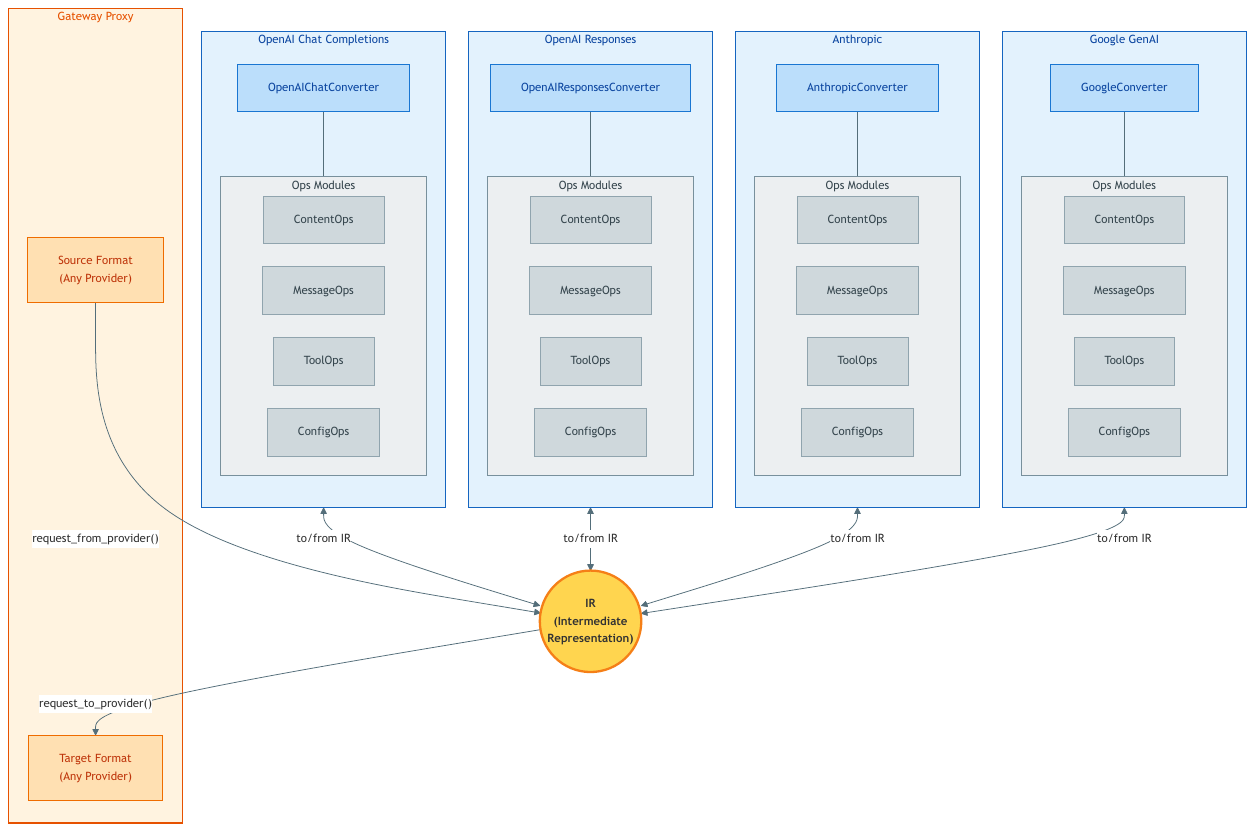}
\caption{Hub-and-spoke architecture of \rosetta. Each provider converter translates bidirectionally between its native format and the \ir. Cross-provider translation composes two converters through the \ir hub.}
\label{fig:architecture}
\end{figure}

\Cref{fig:architecture} illustrates the overall architecture, showing how the hub-and-spoke pattern connects four provider converters through the central \ir.

%% file: sections/implementation.tex
\section{Implementation}
\label{sec:impl}

\rosetta is implemented in Python (approximately 23{,}000 lines of library code, excluding vendored dependencies) and released under the MIT license.
This section describes the key implementation aspects: type system (\cref{sec:impl:types}), converter pipeline (\cref{sec:impl:pipeline}), streaming (\cref{sec:impl:streaming}), provider auto-detection (\cref{sec:impl:detect}), and the gateway proxy (\cref{sec:impl:gateway}).

\subsection{Type System}
\label{sec:impl:types}

The \ir is implemented using Python's \code{TypedDict} with discriminated unions.
Content parts use a \code{type} field as the discriminator:

\begin{lstlisting}[style=pythonstyle,caption={Discriminated union for content parts (simplified).},label={lst:types}]
class TextPart(TypedDict):
    type: Literal["text"]
    text: str

class ToolCallPart(TypedDict):
    type: Literal["tool_call"]
    tool_call_id: str
    tool_name: str
    tool_input: dict[str, Any]
\end{lstlisting}

Role-specific content types constrain which parts may appear in each message role (e.g., \code{ToolCallPart} only in assistant messages), providing static type safety.
Runtime validation functions (\code{validate\_ir\_request}, \code{validate\_ir\_response}) enforce structural invariants.

\subsection{Converter Pipeline}
\label{sec:impl:pipeline}

Each converter exposes six primary entry points:

\begin{itemize}
    \item \code{request\_to\_provider(ir\_request)} $\to$ provider request dict
    \item \code{request\_from\_provider(provider\_request)} $\to$ \code{IRRequest}
    \item \code{response\_to\_provider(ir\_response)} $\to$ provider response dict
    \item \code{response\_from\_provider(provider\_response)} $\to$ \code{IRResponse}
    \item \code{stream\_response\_to\_provider(ir\_events)} $\to$ provider SSE chunks
    \item \code{stream\_response\_from\_provider(chunks)} $\to$ IR stream events
\end{itemize}

Two additional convenience methods (\code{messages\_to\_provider}, \code{messages\_from\_provider}) provide direct message-level conversion without full request wrapping.

A \code{ConversionContext} object threads through the pipeline, accumulating warnings and carrying state between conversion stages.
The context supports two metadata modes:
\begin{itemize}
    \item \textbf{Strip mode} (default): Provider-specific metadata is discarded during conversion, producing clean \ir output suitable for cross-provider forwarding.
    \item \textbf{Preserve mode}: Provider-specific metadata is retained in \code{provider\_metadata} fields, enabling lossless round-trip conversion (A$\to$\ir$\to$A).
\end{itemize}

Internally, each entry point orchestrates the four Ops modules.
For example, \code{request\_from\_provider} proceeds in stages: (1)~ConfigOps extracts generation parameters, (2)~ToolOps extracts tool definitions, (3)~MessageOps (calling ContentOps per part) converts the conversation, and (4)~provider-specific extensions are captured in \code{provider\_extensions}.

\subsection{Streaming}
\label{sec:impl:streaming}

Streaming translation is a key design challenge because providers use fundamentally different Server-Sent Events (SSE)~\citep{sse2015} schemas.
OpenAI Chat emits \code{delta} chunks within a \code{choices[]} array; Anthropic emits typed events (\code{content\_block\_start}, \code{content\_block\_delta}, \code{content\_block\_stop}); Google emits \code{candidates[]} with accumulated parts; and OpenAI Responses emits item-level events.

\rosetta normalizes these into ten \ir stream event types:

\begin{enumerate}
    \item \code{stream\_start}: Session metadata (response ID, model, timestamp).
    \item \code{stream\_end}: End of stream.
    \item \code{content\_block\_start}: Begin a new content block (text, tool call, reasoning).
    \item \code{content\_block\_end}: Finish a content block.
    \item \code{text\_delta}: Incremental text fragment.
    \item \code{reasoning\_delta}: Incremental reasoning/thinking fragment.
    \item \code{tool\_call\_start}: Begin a tool call (ID, name).
    \item \code{tool\_call\_delta}: Incremental tool call arguments (JSON fragment).
    \item \code{finish}: Generation complete, with finish reason.
    \item \code{usage}: Token usage statistics.
\end{enumerate}

A \code{StreamContext} extends \code{ConversionContext} with streaming-specific state: the current block index, a tool-call ID-to-name mapping, accumulated tool-call argument buffers, and deferred payloads for usage and finish events that arrive before the logical end of a content block.
This stateful design ensures correct ordering of events even when providers report information out of sequence (e.g., Google emitting finish reason before the final content delta).

\subsection{Provider Auto-Detection}
\label{sec:impl:detect}

\rosetta includes a heuristic auto-detection module that infers the source provider format from a request body's structure.
The detection examines field presence and types in priority order:

\begin{enumerate}
    \item \textbf{Google GenAI}: Presence of \code{contents} with \code{parts} sub-structure.
    \item \textbf{OpenAI Responses}: Presence of \code{input} or \code{output} with typed items.
    \item \textbf{Anthropic vs.\ OpenAI Chat}: Both use \code{messages}; differentiated by Anthropic's separate \code{system} parameter, \code{anthropic\_version} field, or block-typed content arrays.
\end{enumerate}

Auto-detection enables the gateway proxy to accept requests in any supported format without explicit provider specification.

\subsection{Gateway Proxy}
\label{sec:impl:gateway}

The \rosetta gateway is an HTTP proxy built on Starlette~\citep{starlette2024} that performs live cross-provider translation.
Given a request in format~A destined for a provider expecting format~B, the gateway:

\begin{enumerate}
    \item Auto-detects (or receives as configuration) the source format.
    \item Converts the request: A $\to$ \ir $\to$ B.
    \item Forwards to the upstream provider.
    \item Converts the response: B $\to$ \ir $\to$ A.
    \item Returns the translated response to the client.
\end{enumerate}

For streaming requests, the gateway performs chunk-level SSE translation: each upstream SSE event is converted to an \ir stream event, then re-serialized into the source provider's SSE format, and forwarded to the client in real time.
This ensures that streaming latency overhead is bounded by per-chunk conversion time rather than total response time.

The gateway supports configurable provider endpoints and API keys, making it suitable for local development, testing, and production deployment behind a reverse proxy.

%% file: sections/evaluation.tex
\section{Evaluation}
\label{sec:eval}

We evaluate \rosetta along four dimensions: round-trip fidelity (\cref{sec:eval:fidelity}), streaming correctness (\cref{sec:eval:streaming}), cross-provider translation (\cref{sec:eval:cross}), and conversion performance overhead (\cref{sec:eval:perf}).

\subsection{Evaluation Methodology}

We organize evaluation around the following research questions:
\begin{itemize}
    \item \textbf{RQ1}: Does round-trip conversion (A$\to$\ir$\to$A) preserve all application-relevant fields?
    \item \textbf{RQ2}: Does streaming translation maintain correct event ordering and content integrity?
    \item \textbf{RQ3}: Does cross-provider translation (A$\to$\ir$\to$B) preserve semantic content?
    \item \textbf{RQ4}: What is the latency and throughput overhead of conversion?
\end{itemize}

\subsection{Round-Trip Fidelity (RQ1)}
\label{sec:eval:fidelity}

\subsubsection{Test Design}

We construct a corpus of representative request and response payloads covering:
\begin{itemize}
    \item Simple text conversations (single and multi-turn).
    \item Multi-modal content (text + images, files).
    \item Tool definitions, tool calls, and tool results.
    \item Reasoning/thinking content with signatures.
    \item Complex generation configurations (temperature, top-p, stop sequences, reasoning budgets).
    \item Edge cases: empty content, refusals, citations, multiple choices.
\end{itemize}

For each payload in provider format~A, we perform the round-trip A $\to$ \ir $\to$ A and compare the output against the original using structural equality (ignoring field ordering and whitespace).

\subsubsection{Results}

\begin{table}[t]
\centering
\caption{Unit test counts per conversion category and provider. All tests pass in both strip and preserve metadata modes.}
\label{tab:roundtrip}
\small
\begin{tabular}{@{}lrrrr@{}}
\toprule
\textbf{Category} & \textbf{OpenAI Chat} & \textbf{Anthropic} & \textbf{Google} & \textbf{Responses} \\
\midrule
Content parts      & 22  & 31  & 33  & 43  \\
Messages           & 38  & 23  & 28  & 27  \\
Tool defs/calls    & 25  & 28  & 31  & 46  \\
Config/params      & 23  & 26  & 47  & 44  \\
Full round-trip    & 31  & 41  & 51  & 40  \\
Streaming          & 56  & 70  & 59  & 70  \\
\midrule
\textbf{Total}     & \textbf{199} & \textbf{229} & \textbf{255} & \textbf{304} \\
\bottomrule
\end{tabular}
\end{table}

\Cref{tab:roundtrip} summarizes the test coverage.
The suite contains 987 converter-level unit tests across the four providers, plus 377 additional tests for IR types, base converter logic, auto-detection, and public API surface (1{,}364 total).
In preserve mode, all round-trip tests achieve lossless field-level equality.
In strip mode, provider-specific metadata (e.g., Anthropic's \code{cache\_control}, Google's \code{thought\_signature}) is intentionally discarded, but all semantically meaningful fields are preserved.

\subsubsection{Open Responses Compliance}

\rosetta passes all six tests in the official Open Responses compliance test suite~\citep{openresponses2025}, covering non-streaming text generation, streaming, tool use, multi-turn conversation, image input, and structured output.

\subsection{Streaming Correctness (RQ2)}
\label{sec:eval:streaming}

\subsubsection{Test Design}

We capture real streaming sessions from each provider (using recorded SSE traces) and verify that:
\begin{itemize}
    \item Every upstream event produces the correct sequence of \ir events.
    \item Event ordering is maintained (start before deltas, deltas before end).
    \item Tool call arguments are correctly accumulated across delta events.
    \item Usage and finish events are correctly positioned.
    \item Re-serialization into the source format produces valid SSE.
\end{itemize}

\subsubsection{Results}

The streaming test suite contains 255 test cases (56 OpenAI Chat, 70 Anthropic, 59 Google, 70 Responses; see \cref{tab:roundtrip}).
All four provider converters pass with 100\% event-sequence accuracy.
The \code{StreamContext} correctly handles provider-specific ordering differences:
\begin{itemize}
    \item Anthropic's explicit block lifecycle events map directly to \ir block events.
    \item OpenAI Chat's implicit block boundaries (inferred from delta field presence) are correctly detected.
    \item Google's accumulated-part model (where each chunk contains the full response so far) is correctly differenced to produce incremental deltas.
    \item OpenAI Responses' item-level events are correctly mapped to block-level \ir events.
\end{itemize}

\subsection{Cross-Provider Translation (RQ3)}
\label{sec:eval:cross}

Round-trip fidelity (A$\to$\ir$\to$A) is the easier case because both legs of the conversion share the same provider logic.
Cross-provider translation (A$\to$\ir$\to$B) is the more demanding scenario: it exercises both converters in tandem and exposes semantic gaps between formats.

\subsubsection{Test Design}

We test cross-provider conversion across all six provider pairs (OpenAI Chat$\leftrightarrow$Anthropic, OpenAI Chat$\leftrightarrow$Google, OpenAI Chat$\leftrightarrow$Responses, Anthropic$\leftrightarrow$Google, Anthropic$\leftrightarrow$Responses, Google$\leftrightarrow$Responses) covering:
\begin{itemize}
    \item Simple text conversations (role mapping, content structure).
    \item Multi-modal content (text + inline images).
    \item Tool definitions and tool choice configuration.
    \item Multi-turn conversations with mixed roles.
    \item Bidirectional consistency: A$\to$B$\to$A should recover the semantic content of A.
\end{itemize}

\subsubsection{Results}

All 10 cross-provider conversion tests pass.
Semantic content---message text, roles, tool names, tool parameters, and image data---is preserved across all provider pairs.
The following provider-specific adaptations are correctly handled:
\begin{itemize}
    \item \textbf{Role mapping}: Google's \code{model} role is correctly mapped to/from \code{assistant} in other providers.
    \item \textbf{System prompt location}: Anthropic's top-level \code{system} parameter is correctly extracted from or merged into the message array used by other providers.
    \item \textbf{Content structure}: OpenAI Chat's string-or-array content, Anthropic's block arrays, Google's \code{parts} nesting, and Responses' typed items are all interconvertible.
    \item \textbf{Tool definitions}: Function schema and tool choice configurations translate correctly despite differing nesting structures.
\end{itemize}

The primary limitation of cross-provider translation is the expected loss of provider-specific features that have no equivalent in the target format.
For example, Anthropic's \code{cache\_control} annotations are not representable in Google's format, and Google's \code{grounding\_metadata} has no Anthropic counterpart.
These features are silently dropped during cross-provider conversion (with warnings recorded in the \code{ConversionContext}), while all semantically shared fields are preserved.

\subsection{Performance Overhead (RQ4)}
\label{sec:eval:perf}

\subsubsection{Test Design}

We measure conversion latency using microbenchmarks on representative payloads of varying complexity (simple text, multi-turn with tools, multi-modal).
Each benchmark performs 1{,}000 iterations of the full round-trip conversion (provider$\to$\ir$\to$provider) using Python's \code{time.perf\_counter\_ns()} for nanosecond-resolution timing.
Benchmarks were run on a single core of an Intel Core Ultra 7 155H (32\,GB RAM) with CPython 3.10 on Linux 6.19.

\subsubsection{Results}

\begin{table}[t]
\centering
\caption{Round-trip conversion overhead in microseconds (median over 1{,}000 iterations). ``Req'' denotes request round-trip (provider$\to$\ir$\to$provider); ``Resp'' denotes response round-trip.}
\label{tab:perf}
\small
\begin{tabular}{@{}lrrrr@{}}
\toprule
\textbf{Payload} & \textbf{OpenAI Chat} & \textbf{Anthropic} & \textbf{Google} & \textbf{Responses} \\
\midrule
Simple text (req)    & 21  & 24  & 24  & 22  \\
Multi-turn (req)     & 71  & 75  & 77  & 73  \\
Tool calls (req)     & 44  & 46  & 46  & 44  \\
Simple text (resp)   & 30  & 29  & 31  & 31  \\
Tool calls (resp)    & 55  & 47  & 48  & 57  \\
\bottomrule
\end{tabular}
\end{table}

\Cref{tab:perf} reports round-trip (A$\to$\ir$\to$A) conversion latency across five payload types.
All conversions complete in under 80\,\textmu s at the median, with simple requests taking 21--24\,\textmu s and the most complex multi-turn payloads reaching 71--77\,\textmu s.
At the 95th percentile (P95), latencies remain below 115\,\textmu s for all payload types, with P95 values typically 5--15\% above the median.
These overheads are negligible compared to network round-trip times (typically 50--500\,ms) and model inference latency (100\,ms--10\,s+), representing less than 0.01\% of end-to-end request latency in practice.
Notably, conversion time scales with message count (multi-turn payloads are ${\sim}3\times$ slower than simple text) but is largely uniform across providers, confirming that the Ops-composition architecture does not introduce provider-specific bottlenecks.

\subsubsection{Comparison with LiteLLM}

To contextualize \rosetta's overhead, we benchmark LiteLLM's one-directional \code{transform\_request} (OpenAI Chat$\to$Anthropic, v1.83) against \rosetta's two-hop cross-provider path (OpenAI Chat$\to$\ir$\to$Anthropic) using the same payloads.

\begin{table}[t]
\centering
\caption{Cross-provider conversion latency: LiteLLM (one-directional, OpenAI Chat$\to$Anthropic) vs.\ \rosetta (two-hop, OpenAI Chat$\to$\ir$\to$Anthropic). Median over 1{,}000 iterations; LiteLLM v1.83.}
\label{tab:litellm}
\small
\begin{tabular}{@{}lrrr@{}}
\toprule
\textbf{Payload} & \textbf{LiteLLM (\textmu s)} & \textbf{\rosetta (\textmu s)} & \textbf{Ratio} \\
\midrule
Simple text  & 28  & 24  & 0.8$\times$ \\
Multi-turn   & 34  & 76  & 2.2$\times$ \\
Tool calls   & 29  & 46  & 1.6$\times$ \\
\bottomrule
\end{tabular}
\end{table}

As shown in \cref{tab:litellm}, \rosetta's two-hop conversion is competitive with LiteLLM's single-pass approach: for simple text payloads, \rosetta is actually \emph{faster} (24 vs.\ 28\,\textmu s), while multi-turn and tool-call payloads incur a 1.6--2.2$\times$ overhead.
This modest gap reflects the additional work of constructing typed IR objects at the intermediate hop.
Crucially, both tools operate in the sub-100\,\textmu s range, so the absolute difference is negligible relative to network round-trip times and model inference latency.
The two-hop cost buys bidirectionality, provider neutrality, and lossless round-trip capability---features that LiteLLM does not support.

\subsection{Threats to Validity}
\label{sec:eval:threats}

\paragraph{Construct validity.}
Our fidelity evaluation relies on unit tests authored alongside the converters, which risks circular validation: tests may reflect the implementation's assumptions rather than an independent specification of correct behavior.
We mitigate this in two ways.
First, test payloads are constructed from official provider documentation and real API responses, not generated from the converter code.
Second, \rosetta passes all six tests in the independently maintained Open Responses compliance suite~\citep{openresponses2025}, providing external validation of the core translation logic.

\paragraph{Internal validity.}
The benchmark payloads (simple text, multi-turn, tool calls) are synthetic but representative.
We additionally validated performance on anonymized production payloads from Argo-Proxy (64- and 218-message conversations with 41 tool definitions), confirming that the latency characteristics hold at production scale.
All benchmarks use CPython 3.10 on a single hardware configuration; results may differ on other Python implementations or hardware.

\paragraph{External validity.}
\rosetta currently supports four API standards.
While these cover the dominant LLM API providers by market share, our results do not guarantee that the \ir design or the Ops-composition pattern will generalize equally well to all future providers.
The cross-provider evaluation covers all six bidirectional provider pairs but with a limited number of test cases per pair (10 total); more extensive cross-provider testing would strengthen confidence in translation correctness.

%% file: sections/discussion.tex
\section{Discussion}
\label{sec:discussion}

\subsection{Semantic vs.\ Syntactic Translation}

\rosetta performs \emph{semantic} translation: it maps between provider formats based on the meaning of fields, not their surface syntax.
For example, Anthropic's \code{thinking} content blocks and Google's \code{thought} parts both represent chain-of-thought reasoning and are mapped to the same \ir \code{ReasoningPart}, despite having different JSON structures.
This semantic approach enables cross-provider translation (A$\to$\ir$\to$B) in addition to round-trip conversion.

However, semantic translation inevitably involves judgment calls about equivalence.
When provider formats diverge in expressiveness (e.g., one supports \code{top\_k} and another does not), the \ir takes the union of features, and converters for less expressive providers simply ignore unsupported fields with a warning.
This design choice prioritizes completeness over strict compatibility.

\subsection{Metadata Preservation and Round-Trip Fidelity}

The dual-mode metadata system (strip vs.\ preserve) provides flexibility for different use cases.
Preserve mode enables lossless A$\to$\ir$\to$A round-trips by storing provider-specific attributes in \code{provider\_metadata} fields.
This is essential for testing and debugging converters, and for scenarios where a request must pass through the \ir and return to its original format.

Strip mode provides clean, provider-neutral \ir output suitable for cross-provider forwarding.
In this mode, provider-specific metadata is intentionally discarded, which means that A$\to$\ir$\to$B$\to$\ir$\to$A may not reproduce the original A exactly---but the semantic content is preserved.

\subsection{Limitations}

\paragraph{Coverage.}
\rosetta currently supports four API standards (OpenAI Chat Completions, OpenAI Responses, Anthropic Messages, Google Generative AI), which cover the majority of commercial LLM providers---most emerging providers (Cohere, Mistral, xAI, DeepSeek, etc.) adopt one of these wire formats.
Non-chat modalities (embeddings, fine-tuning, batch processing) are not yet covered.

\paragraph{API evolution.}
LLM APIs evolve rapidly.
When a provider adds new fields or changes semantics, the corresponding converter must be updated.
The modular Ops design localizes these changes (e.g., a new content type only affects ContentOps), but ongoing maintenance is unavoidable.

\paragraph{Semantic gaps.}
Some provider features have no equivalent in other formats.
For example, Anthropic's prompt caching with explicit cache breakpoints has no counterpart in Google's API.
\rosetta can preserve such features in \code{provider\_metadata} for round-trips, but cross-provider translation necessarily drops them.

\paragraph{Performance at scale.}
The translation layer itself adds sub-100\,\textmu s overhead per conversion at the median (see \cref{sec:eval:perf}), which is negligible compared to network and inference latency.
This makes the converter library suitable for direct integration into production applications---for example, embedded in an API gateway, an orchestration framework, or a multi-provider SDK.
The reference gateway included in \rosetta is one such deployment example; it introduces an additional network hop and serialization cycle, which may matter in latency-sensitive settings.
In such cases, applications can invoke the translation layer in-process to avoid the extra hop entirely.

\subsection{Deployment Experience}

\rosetta serves as the translation layer for \textsc{Argo-Proxy}~\citep{argoproxy2024}, an LLM API gateway deployed at Argonne National Laboratory.
Argo-Proxy provides researchers on the Argonne network with unified access to multiple LLM providers (OpenAI, Anthropic, Google) through a single OpenAI-compatible endpoint.
Its previous architecture (v2.x) relied on hand-written format-mapping code that only covered the OpenAI Chat Completions format, and could not keep pace with the evolving demands of Anthropic's Messages API and the newer OpenAI Responses API.
The third-generation architecture (v3.0, currently in beta after 13 pre-release iterations) replaced this bespoke translation layer with \rosetta's converter library, eliminating approximately 2{,}000 lines of ad hoc mapping code while gaining support for all four API standards and bidirectional streaming.

The integration exercises \rosetta's core capabilities in a production setting: request translation (OpenAI Chat $\to$ \ir $\to$ target provider), response translation (provider $\to$ \ir $\to$ OpenAI Chat), and streaming event normalization across all supported providers.
The 13-iteration beta cycle (v3.0.0b1--b13) surfaced several edge cases---notably, inconsistent streaming event ordering across providers and corner cases in tool-call argument accumulation---that led to improvements in the converter library itself.
This feedback loop between a production deployment and the library's test suite provides a form of real-world validation beyond unit tests alone.

\subsection{Future Directions}

\paragraph{Provider coverage.}
Since most emerging LLM providers adopt one of the four supported API standards (most commonly OpenAI Chat Completions), they can already be served by the existing converters.
For providers with minor deviations from a standard format, we plan to support configurable \emph{provider adaptors} that map provider-specific endpoints, authentication, and field variations onto an existing converter, reducing per-provider effort to configuration rather than code.

\paragraph{Conformance testing.}
\rosetta currently passes the Open Responses compliance suite (\cref{sec:eval:fidelity}), but no analogous third-party suite exists for the other three provider formats.
Developing a comprehensive, independently maintained conformance test corpus---ideally derived from real API traffic---would strengthen validation and help track correctness as both \rosetta and provider APIs evolve.

\paragraph{Schema evolution.}
As the LLM ecosystem matures, new content types (video, structured data), interaction patterns (multi-agent, agentic workflows), and capabilities (real-time voice, computer use) will require \ir extensions.
The \code{provider\_extensions} mechanism provides an escape hatch, but frequently used extensions should be promoted to first-class \ir fields.

%% file: sections/conclusion.tex
\section{Conclusion}
\label{sec:conclusion}

The central finding of this work is that despite substantial surface-level divergence, the four major LLM API providers share a common semantic core---role-tagged messages, typed content parts, tool definitions with JSON Schema inputs, and incremental streaming events---that can be captured by a compact, provider-neutral IR (9 content-part types, 10 stream event types).
The practical challenge is not deep semantic incompatibility but the \emph{combinatorial surface} of syntactic variations: each provider makes different choices about field naming, nesting depth, content encoding, role vocabulary, and streaming granularity, and these differences multiply across feature dimensions (content, tools, config, streaming).
A hub-and-spoke IR is effective precisely \emph{because} the divergence is syntactic: a shared semantic core makes faithful translation feasible, while the combinatorial cost of pairwise adaptation makes an intermediate representation worthwhile.

\rosetta demonstrates this through 1{,}364 passing tests---including the Open Responses compliance suite---lossless round-trip fidelity in preserve mode, and sub-millisecond conversion overhead.
The Ops-composition architecture confines provider-specific complexity to well-bounded modules, and the library's deployment in Argo-Proxy at Argonne National Laboratory validates its production readiness.

The primary open challenge is coverage breadth: the four supported API standards cover most commercial providers, but non-chat modalities (embeddings, fine-tuning, batch processing) are not yet addressed.
As the ecosystem continues to grow, we expect the IR's union-based design to accommodate new providers with incremental effort, while the \code{provider\_extensions} mechanism provides a pragmatic escape hatch for features that resist standardization.
\rosetta is available at \url{https://github.com/Oaklight/llm-rosetta}.